\documentclass[twocolumn,prb]{revtex4}
\usepackage{graphicx}
\usepackage{dcolumn}
\usepackage{bm}
\usepackage{amsmath}
\usepackage{amsthm}
\usepackage{amsfonts}

\def\reals{{\mathbb R}}

\begin{document}

\preprint{}
\title{Theory of  Fulde-Ferrell-Larkin-Ovchinnikov state of  superconductors 
with and without inversion symmetry: Hubbard model approach}
\author{Takehito Yokoyama, Seiichiro Onari, and Yukio Tanaka }
\affiliation{Department of Applied Physics, Nagoya University, Nagoya, 464-8603, Japan%
\\
and CREST, Japan Science and Technology Corporation (JST) Nagoya, 464-8603,
Japan }
\date{\today}

\begin{abstract}
 We study Fulde-Ferrell-Larkin-Ovchinnikov (FFLO) state  of superconductors with and without inversion symmetry based on the  Hubbard model on the square  lattice near half-filling, using the random phase approximation. We show that center of mass momentum 
$Q$ tends to be parallel to $x$- or $y$-axis in the presence of  inversion symmetry, while $Q$ vector is likely to be perpendicular  to the magnetic field in the absence of  inversion symmetry. We also clarify that $d+f$-wave pairing is favored and the hetero spin triplet $f$-wave state is present in the FFLO state unlike state in the superconductors only with the Rashba type spin-orbit coupling (RSOC) originating from the broken inversion symmetry. The triplet $f$-wave state is enhanced by magnetic field and the RSOC. This stems from the reduction of the spin susceptibilities by the magnetic field and the RSOC. 
\end{abstract}

\maketitle




%

%




\section{Introduction}
Fulde-Ferrell-Larkin-Ovchinnikov (FFLO) state was predicted more than fourty years ago\cite{Ful64,Lar64} and has been intensively studied.\cite{Cas04,Matsuda}  FFLO state is characterized by the formation of Cooper pairing in a magnetic field with nonzero center of mass momentum $Q$: $(k \uparrow,  -k+Q \downarrow)$. Up to now, various properties of the FFLO state have been predicted.\cite{Shimahara2,Shimahara4,Maki,Shimahara5,Adachi,Agterberg2,Won,Buzdin2,Vorontsov,Tanaka2007} Recently, a strong candidate of the FFLO state, CeCoIn$_5$, has been discovered.\cite{Rad03,Bia03,Wat04,Wat04a,Cap04,Cor06,Kak05,Kum06,Mic06,Gra06,Mit06}
 This material meets necessary conditions for the existence of FFLO state, namely being in the clean limit,\cite{Movshovich} two dimensional electric nature,\cite{Hall}  paramagnetically limited upper critical field,\cite{Bianchi}  and $d$-wave superconductivity\cite{Izawa,Aoki,Vorontsov2}, and is also known as a strongly correlated system.  However, all previous studies of FFLO state did not take into account the electron-electron repulsion beyond Fermi liquid corrections. \cite{Bur94,Shi94,Tak69,Vor06} Moreover, most of the studies of pairing symmetry of FFLO state are based on the variational method. 

Another aspect of the FFLO state which has not been so emphasized is that since magnetic field breaks time reversal symmetry, singlet and triplet pairings should be mixed in the FFLO state according to the Pauli's principle, as in a ferromagnet attached to a singlet superconductor (FFLO-like state is expected to appear in the ferromagnet in ferromagnet/superconductor junctions).\cite{Buzdin,Bergeret,Kadigrobov,Braude,Yokoyama,Asano2007} Actually, it is predicted that FFLO state becomes more stable by mixing $p$-wave pairing.\cite{Matsuo}

Recently, it has been found that FFLO-like state can also appear  in surface superconductivity or noncentrosymmetric superconductors.\cite{Barzykin,Dimitrova,Kaur,Agterberg,Tanaka}  Starting from the discovery of  heavy fermion superconductor ${\rm CePt_{3}Si}$, \cite{Bauer} the study of noncentrosymmetric superconductors has recenlty become a hot topic in condensed matter physics. \cite{Akazawa,Kimura,Sugitani,Frigeri,Yogi,Izawa2,Samokhin,Bauer2,Sergienko,Mineev,Bonalde,Fujimoto,Fujimoto2,Fujimoto3,Fujimoto4,Hayashi,Yokoyama2,Yokoyama3,Iniotakis,Yanase,Ben,Samokhin2}
 Due to the  broken inversion symmetry, 
Rashba type spin-orbit coupling (RSOC) is induced in the noncentrosymmetric superconductors,\cite{Rashba,Edelstein}  and therefore  spin-singlet pairing and spin-triplet pairing can be mixed in superconducting state. \cite{Gor'kov} However, the study of FFLO state in noncentrosymmetric superconductors is insufficient, especially from the viewpoint of strongly correlated systems. 

In this paper, we study FFLO state  of superconductors with and without inversion symmetry based on the  Hubbard model on the square  lattice near half-filling, using the random phase approximation (RPA).  By solving the linearized $\acute{{\rm E}}$liashberg's equations directly, we can elude the difficulty of variational method. We show that center of mass momentum 
$Q$ tends to be parallel to $x$- or $y$-axis in the presence of  inversion symmetry, while the $Q$ vector tends to be perpendicular  to the magnetic field in the absence of  inversion symmetry. We also clarify that $d+f$-wave pairing is favored and the hetero spin triplet $f$-wave state is present in the FFLO state in constrast to the superconductors only with RSOC. The triplet $f$-wave state is enhanced by magnetic field and the RSOC which stems from the reduction of the spin susceptibilities by the magnetic field and the  RSOC. 

The organization of the present paper is as follows. 
In section II, we explain our model and introduce $\acute{{\rm E}}$liashberg's equations with the RPA. 
In section  III, we present calculated results of  the eigenvalue of $\acute{{\rm E}}$liashberg's equations and the gap functions. 
In section IV, a summary of the results in the present paper is given.

\section{Formulation}
We consider  the square lattice without inversion center in a magnetic field oriented to the $x$-axis. The  Hubbard model can be written as 
\begin{eqnarray}
 H =  - \sum\limits_{k,\sigma} {\left( {2t\left( {\cos k_x  + \cos k_y } \right) + \mu } \right)c_{k\sigma }^\dag  } c_{k\sigma } \nonumber  \\  +  \sum\limits_{k,s,s' }\left( {\begin{array}{*{20}c}   {h+ \lambda \sin k_y }  \\
   { - \lambda \sin k_x }  \\
   0  \\
\end{array}} \right) \cdot {\bm{\sigma }}_{s,s' }  {c_{ks}^\dag  } c_{k s' } 
  + U\sum\limits_k {n_{k \uparrow } } n_{k \downarrow } .
\end{eqnarray}
 Here, $k$ represents two dimensional vector. We set lattice constant to be unity. The first term is the dispersion relation  and the second term consists of the Zeeman term with the energy $h$ and the RSOC with coupling constant $\lambda$. The third  term represents on-site  electron-electron repulsion.

Then, the bare Green's functions have the following form in the 2$\times$2 spin space:
\begin{widetext}
\begin{eqnarray}
 \left( {\begin{array}{*{20}c}
   {G_{ \uparrow  \uparrow } \left( {k,i\omega _n } \right)} & {G_{ \uparrow  \downarrow } \left( {k,i\omega _n } \right)}  \\
   {G_{ \downarrow  \uparrow } \left( {k,i\omega _n } \right)} & {G_{ \downarrow  \downarrow } \left( {k,i\omega _n } \right)}  \\
\end{array}} \right)
 = G_ +  \left( {k,i\omega _n } \right) 
  + \frac{1}{{\sqrt {(\lambda \sin k_x )^2  + (\lambda \sin k_y  + h)^2 } }}\left( {\begin{array}{*{20}c}
   {\lambda \sin k_y  + h}  \\
   { - \lambda \sin k_x }  \\
   0  \\
\end{array}} \right) \cdot {\bm{\sigma }}G_ -  \left( {k,i\omega _n } \right), \\ 
 G_ \pm  \left( {k,i\omega _n } \right) = \frac{1}{2}\left( {\frac{1}{{i\omega _n  - \xi _ +  }} \pm \frac{1}{{i\omega _n  - \xi _ -  }}} \right), \\ 
 \xi _ \pm   =  - 2t\left( {\cos k_x  + \cos k_y } \right) - \mu  
 \pm \sqrt {(\lambda \sin k_x )^2  + (\lambda \sin k_y  + h)^2 }   \label{xi}
\end{eqnarray}
\end{widetext}
with Matsubara frequency $\omega _n$.

\begin{figure}[htb]
\begin{center}
\scalebox{0.4}{
\includegraphics[width=22.0cm,clip]{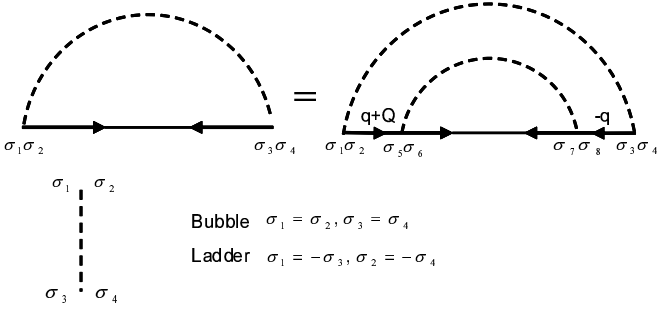}}
\end{center}
\caption{ The diagrammatic representation of the linearized $\acute{{\rm E}}$liashberg's equations. Broken line and $\sigma_i (i=1-8)$ represent effective interaction and spin indices, respectively. In the RPA, we take into account the contributions from bubble and ladder types of diagrams, which impose some relations on spin indices as shown in this figure.}
\label{f1}
\end{figure}

The linearized $\acute{{\rm E}}$liashberg's equations with RPA in the weak coupling approximation are described as (see also Fig. \ref{f1})
\begin{widetext}
\begin{eqnarray}
 \Lambda \Delta _{ss} \left( k \right) = \frac{1}{{\beta N}}\sum\limits_{q,\omega _m ,\sigma ,\sigma ' } (G_{s\sigma } \left( {q+Q,i\omega _m } \right)G_{s\sigma ' } \left( { - q, - i\omega _m } \right)\left( { - \Gamma _{ss} \left( {k - q} \right)} \right)
\nonumber \\
 + G_{ - s\sigma } \left( {q+Q,i\omega _m } \right)G_{ - s\sigma ' } \left( { - q, - i\omega _m } \right) U^2 \chi _{lad}^{ s,-s } \left( {k + q} \right))\Delta _{\sigma \sigma ' } \left( q \right),  \\
 \Lambda \Delta _{s, - s} \left( k \right) = \frac{1}{{\beta N}}\sum\limits_{q,\omega _m ,\sigma ,\sigma ' } {(G_{s\sigma } \left( {q+Q,i\omega _m } \right)G_{ - s\sigma ' } \left( { - q, - i\omega _m } \right)\left( { - \Gamma _{s, - s} \left( {k - q} \right) + U^2 \chi _{lad}^{ss} \left( {k + q} \right)} \right))} \Delta _{\sigma \sigma ' } \left( q \right), \\ 
  - \Gamma _{s ,s } \left( k \right) =   \frac{1}{2} U^2 \chi _C \left( k \right) + \frac{1}{2}U^2 \chi _S \left( k \right), \label{gammauu} \\ 
  - \Gamma _{s , - s } \left( k \right) =  - U  + \frac{1}{2} U^2 \chi _C \left( k \right) - \frac{1}{2} U^2 \chi _S \left( k \right) - U^2 {\mathop{\rm Im}\nolimits} \chi ^{s , -s } \left( k \right) \label{gamma}
\end{eqnarray}
\end{widetext}
with $s=\uparrow,  \downarrow$, center of mass momentum $Q$, and inverse temperature $\beta$. 
Here, $\chi _S$ and $\chi _C$ are spin and charge susceptibilities at $\omega _n=0$, respectively, which are obtained by  
 $\chi _S  = \chi ^{ \uparrow  \uparrow }  - {\mathop{\rm Re}\nolimits} \chi ^{ \uparrow  \downarrow }$ and 
$ \chi _C  = \chi ^{ \uparrow  \uparrow }  + {\mathop{\rm Re}\nolimits} \chi ^{ \uparrow  \downarrow }  $. Note that 
 $\chi ^{ \uparrow  \uparrow }  = \chi ^{ \downarrow  \downarrow }$ and 
 $\chi ^{ \uparrow  \downarrow }  = \left( {\chi ^{ \downarrow  \uparrow } } \right)^*$ are satisfied. 
$\chi ^{ \uparrow  \uparrow }$ and $\chi ^{ \uparrow  \downarrow }$ are given by 
\begin{eqnarray}
 \left( {\begin{array}{*{20}c}
   {\chi ^{ \uparrow  \uparrow } \left( {k } \right)}  \\
   {\chi ^{ \uparrow  \downarrow } \left( {k } \right)}  \\
\end{array}} \right) = \frac{1}{A}\left( {\begin{array}{*{20}c}
   {\chi _1 }   \\
   {\chi _2  - U \left( {\chi _1 ^2  - \left| {\chi _2 } \right|^2 } \right)}  \\
\end{array}} \right), \\
A = 1 - U^2 \left( {\chi _1 ^2  - \left| {\chi _2 } \right|^2 } \right) 
+ 2U{\mathop{\rm Re}\nolimits} \chi _2 ,
\end{eqnarray}
\begin{eqnarray}
 \chi _1 (k)=  - \frac{1}{{\beta N}}\sum\limits_{q,\omega _n} {G_ +  \left( {k + q,i\omega _n  } \right)G_ +  \left( {q,i\omega _n } \right)},  \\ 
 \chi _2 (k) =  - \frac{1}{{\beta N}}\sum\limits_{q,\omega _n } {G_{ \uparrow  \downarrow } \left( {k + q,i\omega _n } \right)G_{ \downarrow  \uparrow } \left( {q,i\omega _n } \right)}  .
\end{eqnarray}
$ \chi _{lad}^{ \uparrow  \uparrow }$ and $\chi _{lad}^{ \uparrow  \downarrow } $ are defined as 
\begin{eqnarray}
\left( {\begin{array}{*{20}c}
   \chi _{lad}^{ \uparrow  \uparrow } (k)  \\
   \chi _{lad}^{ \uparrow  \downarrow } (k)  \\
\end{array}} \right) = \frac{1}{B}\left( {\begin{array}{*{20}c}
   { - \chi _1  + U\left( {\chi _1 ^2  - \left| {\chi _2 } \right|^2 } \right)}  \\
   { - \chi _2 }  \\
\end{array}} \right),
\end{eqnarray}
$B=\left( {1 - U\chi _1 } \right)^2  - \left| {U\chi _2 } \right|^2 $.
Notice that $\chi _{lad}^{ \uparrow  \uparrow }  = \chi _{lad}^{ \downarrow  \downarrow }$ and $  \chi _{lad}^{ \uparrow  \downarrow }  = \left( {\chi _{lad}^{ \downarrow  \uparrow } } \right)^* $ are satisfied.

\begin{figure}[htb]
\begin{center}
\scalebox{0.4}{
\includegraphics[width=18.0cm,clip]{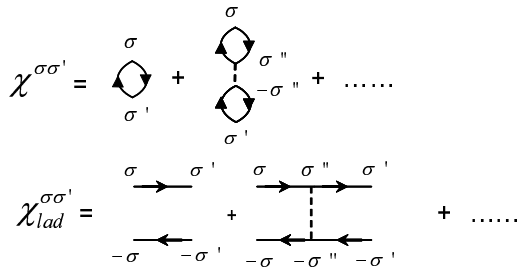}}
\end{center}
\caption{ The diagrammatic representation of $\chi^{ \sigma  \sigma' }$ and $\chi _{lad}^{ \sigma  \sigma' }$. The summation over the internal spin indices, e.g.,  $\sigma''$ has to be taken. }
\label{f15}
\end{figure}

In the RPA, we take into account the contributions from bubble and ladder types of diagrams. \cite{Shimahara3} In the above, $\chi ^{ \uparrow  \uparrow }$ and $\chi ^{ \uparrow  \downarrow }$ stem from the bubble type of diagrams, while $\chi _{lad}^{ \uparrow  \uparrow }$  and $\chi _{lad}^{ \uparrow  \downarrow }$  originate from the ladder type of diagrams (see Fig. \ref{f15}). 
Within the RPA,  self-energy corrections and the frequency dependence 
of effective interaction are ignored. 
However, we can grasp the essence of the physics by the RPA.  
\cite{Miyake,Scalapino}

By solving the $\acute{{\rm E}}$liashberg's equations, we can obtain the 
gap functions. 
We define singlet  component of the pair potential  and triplet one with $S_z=0$ for a later convenience as
\begin{eqnarray}
 \Delta _s  = (\Delta _{ \uparrow  \downarrow }  - \Delta _{ \downarrow  \uparrow } )/2, \\ 
 \Delta _t  = (\Delta _{ \uparrow  \downarrow }  + \Delta _{ \downarrow  \uparrow } )/2 .
\end{eqnarray}

\section{Results}
In the following, we set $t=1$ and parameters as $T=0.01, N=256\times256, U=1.7$ and, $n=0.9$. Here, $T, N,$ and $n$ denote the temperature, k-point meshes and the band filling, respectively. For example, the magnitude of $\lambda$ in CePt$_3$Si is $\lambda \sim 0.3$.\cite{Samokhin}

\subsection{Center of mass momentum}

\begin{figure}[htb]
\begin{center}
\scalebox{0.4}{
\includegraphics[width=16.0cm,clip]{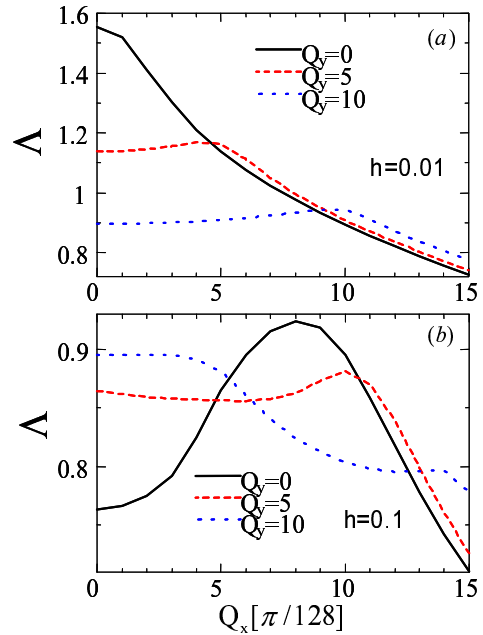}}
\end{center}
\caption{ (color online) Eigenvalue $\Lambda$ as a function of the center of mass momentum at $\lambda=0$. (a) $h=0.01$. (b) $h=0.1$.  }
\label{f2}
\end{figure}

\begin{figure}[htb]
\begin{center}
\scalebox{0.4}{
\includegraphics[width=17.0cm,clip]{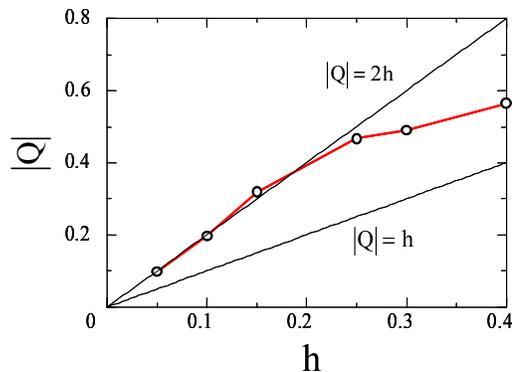}}
\end{center}
\caption{(color online) Magnitude of $Q$ as a function of $h$. $\left| Q \right| = h$ and $\left| Q \right| = 2h$ are depicted for comparison. }
\label{f14}
\end{figure}

\begin{figure}[htb]
\begin{center}
\scalebox{0.4}{
\includegraphics[width=16.0cm,clip]{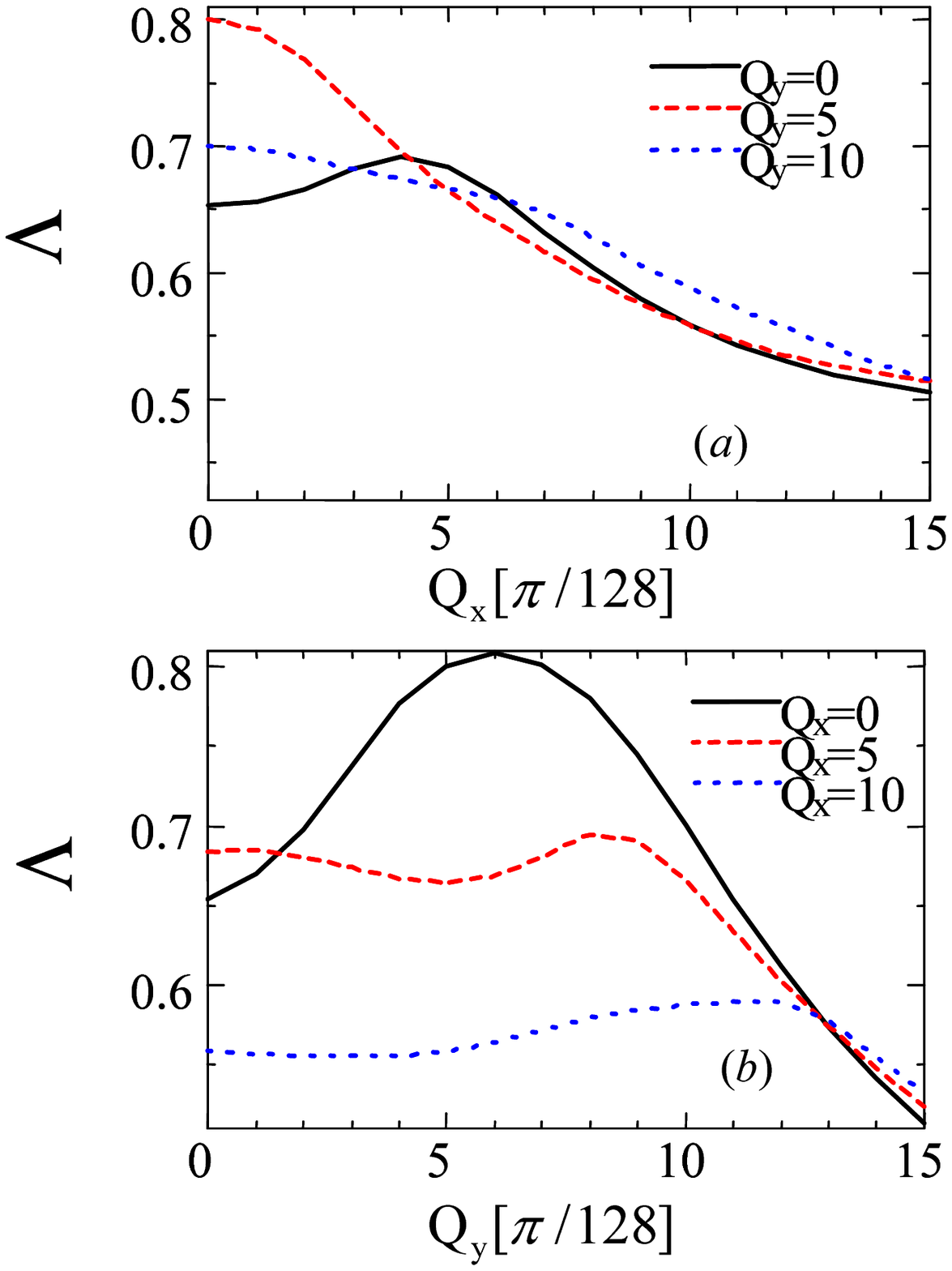}}
\end{center}
\caption{(color online) Eigenvalue $\Lambda$ as a function of the center of mass momentum at $h=0.1$ and $\lambda=0.3$.  }
\label{f3}
\end{figure}

\begin{figure}[tb]
\begin{center}
\scalebox{0.4}{
\includegraphics[width=16.0cm,clip]{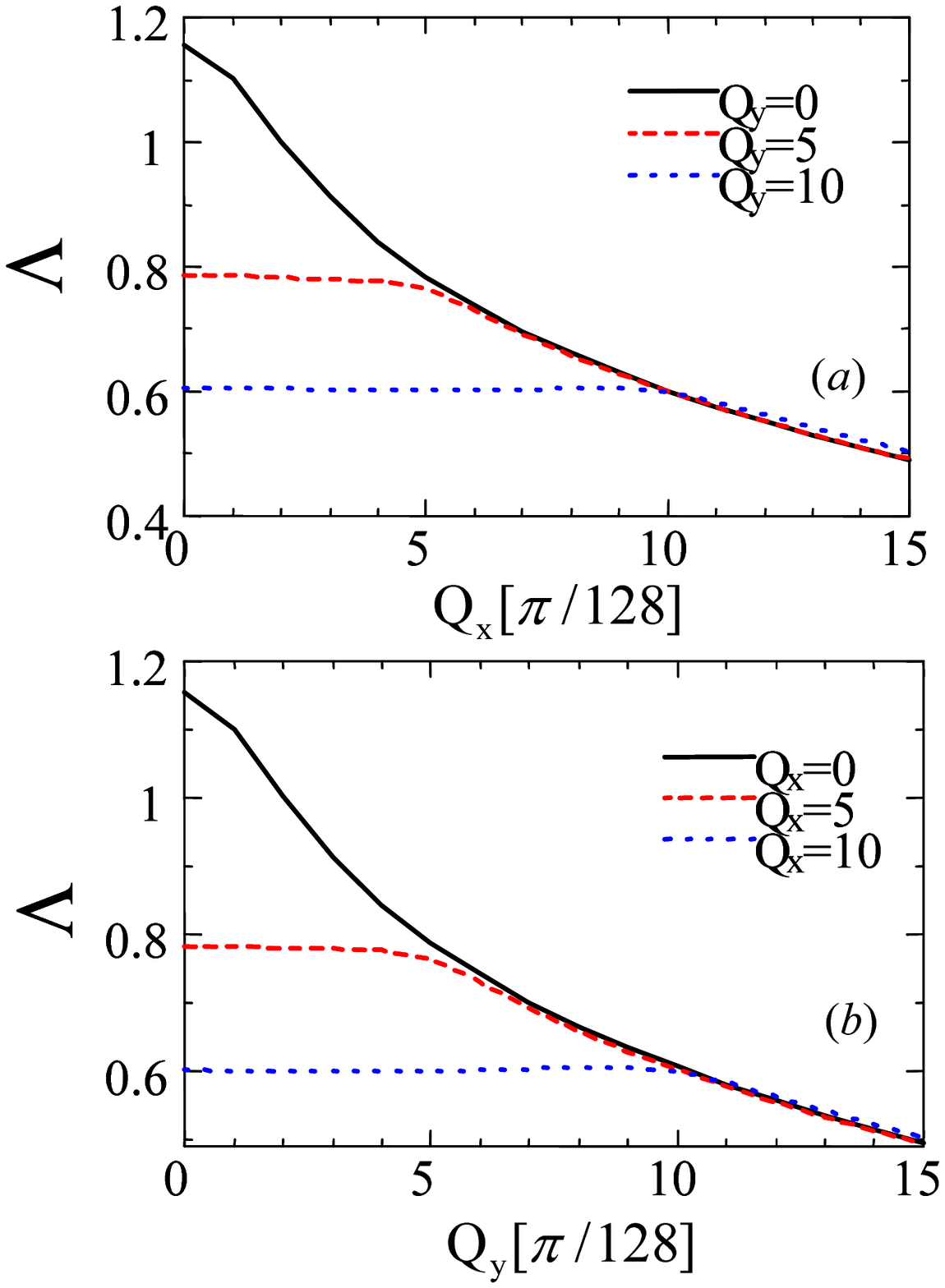}}
\end{center}
\caption{(color online) Eigenvalue $\Lambda$ as a function of the center of mass momentum at $h=0$ and $\lambda=0.3$.}
\label{f4}
\end{figure}

\begin{figure}[tb]
\begin{center}
\scalebox{0.4}{
\includegraphics[width=15.0cm,clip]{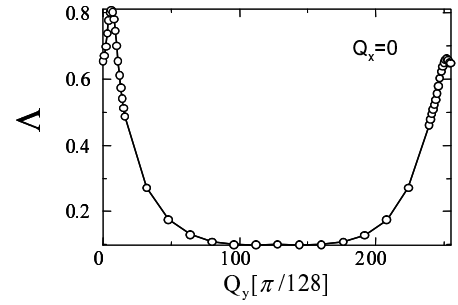}}
\end{center}
\caption{ Eigenvalue $\Lambda$ as a function of the center of mass momentum at $Q_x=0$, $h=0.1$ and $\lambda=0.3$.}
\label{f5}
\end{figure}

\begin{figure}[tb]
\begin{center}
\scalebox{0.4}{
\includegraphics[width=16.0cm,clip]{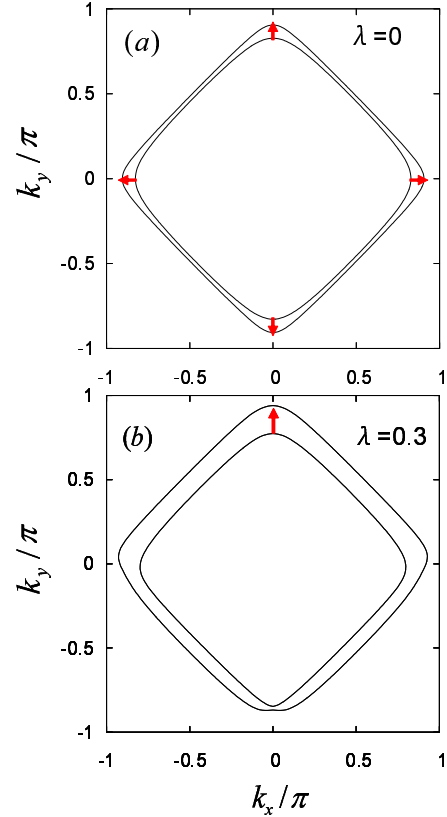}}
\end{center}
\caption{ (color online) Fermi surfaces at $h=0.1$, with (a) $\lambda=0$ and (b) $\lambda=0.3$. Arrows denote nesting vectors. }
\label{f6}
\end{figure}

Let us first discuss the center of mass momentum $Q \equiv (Q_x ,Q_y )$ in unit of $\pi/128$ in order to study the stability of the FFLO state. 
The superconducting transition temperature is determined by the condition $\Lambda=1$. Since $\Lambda$ is a decreasing function of temperature, the state with larger $\Lambda$ is favored. Thus, we study $\Lambda$ as a function of $Q$  to obtain optimal  $Q$. 

Figure \ref{f2} shows the eigenvalue $\Lambda$ as a function of the center of mass momentum at $\lambda=0$. For small magnitude of the magnetic field, $h=0.01$, the state with negligibly small $Q$ is favorable as shown in Fig. \ref{f2} (a).  For a large magnitude of the magnetic field, $h=0.1$, the finite momentum state with $Q_x=8$ and $Q_y=0$  is favorable as shown in Fig. \ref{f2} (b). Therefore, the vector $Q$ tends to be directed to $x$-axis. 
The magnitude of $Q$ is given by $\left| Q \right| \sim h$. In order to study the validity of this relation, we show magnitude of the $Q$ vector as a function of $h$ in Fig. \ref{f14}. The approximate relation $\left| Q \right| \sim h$ holds for wide range of $h$. 
Note that the same result can be obtained by exchanging $Q_x$ by $Q_y$,  $Q_x$ by $-Q_x$, or  $Q_y$ by $-Q_y$ due to the four fold symmetry of the square lattice. Thus, we can conclude that the $Q$ vector is likely to be parallel to $x$- or $y$-axis. \cite{Shimahara4,Shimahara5,Maki,Vorontsov} 

A similar plot in the presence of the RSOC is shown in Fig. \ref{f3}, where we choose $h=0.1$ and $\lambda=0.3$. In this case, the vector $Q$ tends to have a finite value ($Q=(0, 6)$)  and be parallel to $y$-axis. The magnitude of $Q$ is roughly given by $\left| Q \right| \sim h$.  Note that finete momentum  state is unfavorable in the presence of the RSOC but the absence of the magnetic field as shown in  Fig. \ref{f4}, where we choose $h=0$ and $\lambda=0.3$. 
The state with $Q$ and that with $-Q$ are not degenerate  due to the inversion asymmetry as shown in Fig. \ref{f5}, where  $\Lambda$ is plotted for all $Q_y$.

The direction of the vector $Q$ can be understood by the structure of split  Fermi surfaces. Figure \ref{f6} depicts  Fermi surfaces at $h=0.1$, with (a) $\lambda=0$ and (b) $\lambda=0.3$. It is mentioned in Refs.\cite{Shi94,Shimahara2} that the good nesting vector is important for the FFLO state as can be seen from the term $G(q+Q) G(-q)$ in the linearized $\acute{{\rm E}}$liashberg's equations. The summation over $\omega_n$ in $G(q+Q) G(-q)$ leads to the term 
\begin{eqnarray}
\frac{{f(\xi _\sigma  (q + Q)) - f( - \xi _{\sigma '} ( - q))}}{{\xi _\sigma  (q + Q) + \xi _{\sigma '} ( - q)}}
\end{eqnarray}
with $\sigma, \sigma'= \pm$. When this term has a large value, $\xi_\sigma  (q + Q) \xi_{\sigma '} ( - q) >0$ should be satisfied (otherwise the numerator of this term becomes negligible). Therefore, the nesting vector $Q$ should connect nearest Fermi surfaces and  cannot be $\sim (\pm \pi, \pm \pi)$.  Thus, we  see that $Q$ vector tends to be parallel to $x$- or $y$-axis for $\lambda=0$ and to $y$-axis for nonzero $\lambda$ as described by arrows in Fig. \ref{f6}. We can also understand the result for nonzero $\lambda$ as follows. 
 Since larger Fermi surface moves to the van Hove singularity $(0,\pi)$ and the smaller one moves to the opposite direction with the increase of the RSOC  as shown in Fig. \ref{f6}(b), the formation of Cooper pairing on the larger Fermi surface, which has a finite momentum parallel to $y$-axis, contributes dominantly to the superconductivity. Thus,  $Q$ vector is oriented to $y$-axis for nonzero $\lambda$. 
 With a similar analysis,  we can expect that $Q$ vector tends to be perpendicular to the magnetic field in the presence of the RSOC for arbitrary direction of magnetic field. 
 For magnetic field, $h=(h_x, h_y)$, the dispersion relations become
\begin{eqnarray}
\xi _ \pm   =  - 2t\left( {\cos k_x  + \cos k_y } \right) - \mu \nonumber \\ 
 \pm \sqrt {(h_y  - \lambda \sin k_x )^2  + (\lambda \sin k_y  + h_x )^2 } .
\end{eqnarray}
Then, 
\begin{eqnarray}
\left( {\begin{array}{*{20}c}
   {\sin k_x }  \\
   {\sin k_y }  \\
   0  \\
\end{array}} \right) = 1/\lambda \left( {\begin{array}{*{20}c}
   {h_y }  \\
   { - h_x }  \\
   0  \\
\end{array}} \right) = 1/\lambda h \times \left( {\begin{array}{*{20}c}
   0  \\
   0  \\
   1  \\
\end{array}} \right)
\end{eqnarray}
 is satisfied when split Fermi surfaces intersect with each other. This indicates that split Fermi surfaces move to the opposite directions which are perpendicular to the magnetic field. 
 Hence, we see that $Q$ vector tends to be perpendicular to the magnetic field in the presence of the RSOC, as predicted in other works.\cite{Barzykin,Dimitrova,Kaur,Agterberg} 

It should be remarked that $\Lambda$ is reduced as $h$ or $\lambda$ increase. Hence, the superconducting transition temperature is reduced with the increase of them. This can be also predicted by the reduction of spin susceptibilities (see  Fig. \ref{f10} and Fig. \ref{f13}).

\subsection{Pairing symmetry}

\begin{figure}[tb]
\begin{center}
\scalebox{0.4}{
\includegraphics[width=22.0cm,clip]{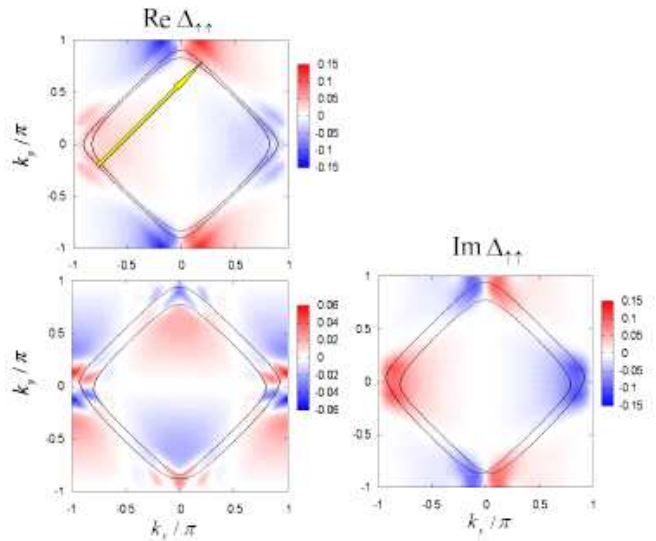}}
\end{center}
\caption{ (color) Real and imaginary parts of gap function $\Delta _{ \uparrow  \uparrow }$ at $h=0.1$.   We take $\lambda=0$, $Q_x=8$ and $Q_y=0$ in the upper figure and $\lambda=0.3$, $Q_x=0$ and $Q_y=6$ in the lower figures. Solid lines represent Fermi surfaces. Yellow arrow represents typical scattering process. }
\label{f7}
\end{figure}

\begin{figure}[tb]
\begin{center}
\scalebox{0.4}{
\includegraphics[width=17.0cm,clip]{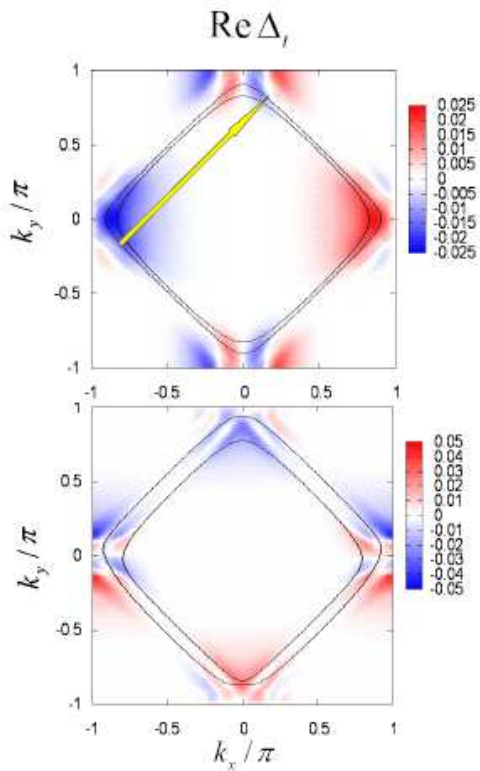}}
\end{center}
\caption{ (color) Real part of triplet gap function $\Delta _t$.  We take $\lambda=0$, $Q_x=8$ and $Q_y=0$ in the upper figure and $\lambda=0.3$, $Q_x=0$ and $Q_y=6$ in the lower figure. Solid lines represent Fermi surfaces.  Yellow arrow represents typical scattering process.}
\label{f8}
\end{figure}

\begin{figure}[tb]
\begin{center}
\scalebox{0.4}{
\includegraphics[width=17.0cm,clip]{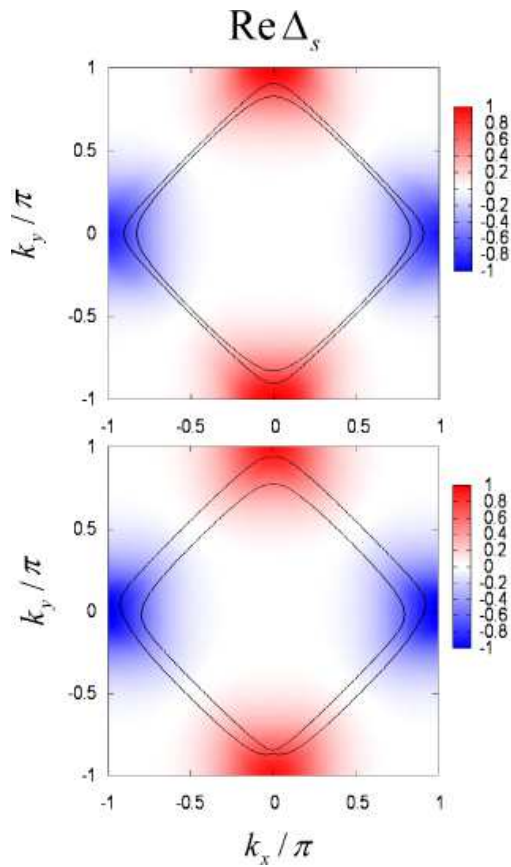}}
\end{center}
\caption{ (color) Real part of singlet gap function $\Delta _s$. $\lambda=0$, $Q_x=8$ and $Q_y=0$ in the upper figure and $\lambda=0.3$, $Q_x=0$ and $Q_y=6$ in the lower figure. Solid lines represent Fermi surfaces. }
\label{f9}
\end{figure}

\begin{figure}[tb]
\begin{center}
\scalebox{0.4}{
\includegraphics[width=18.0cm,clip]{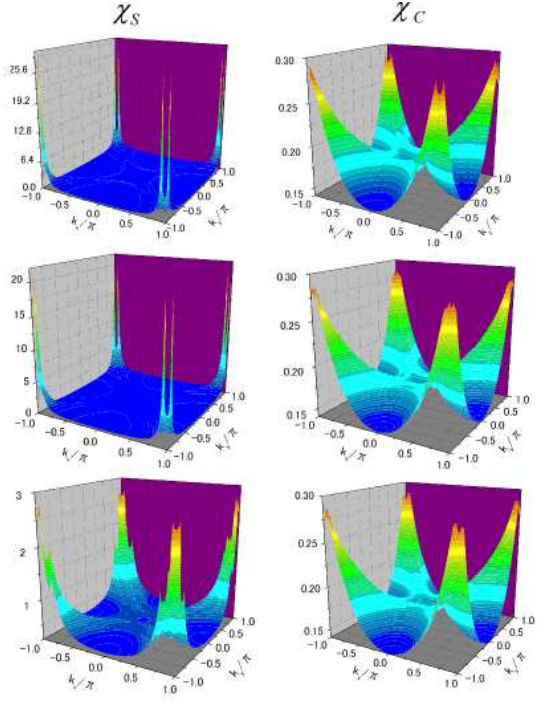}}
\end{center}
\caption{ (color online) Spin and charge susceptibilities ($\chi _{S}$ and $\chi _{C}$).  We set  $h=0.01$ and $\lambda=0$ in the upper figures, $h=0.1$ and $\lambda=0$ in the middle figures, and $h=0.1$ and $\lambda=0.3$ in the lower figures.}
\label{f10}
\end{figure}

\begin{figure}[tb]
\begin{center}
\scalebox{0.4}{
\includegraphics[width=18.0cm,clip]{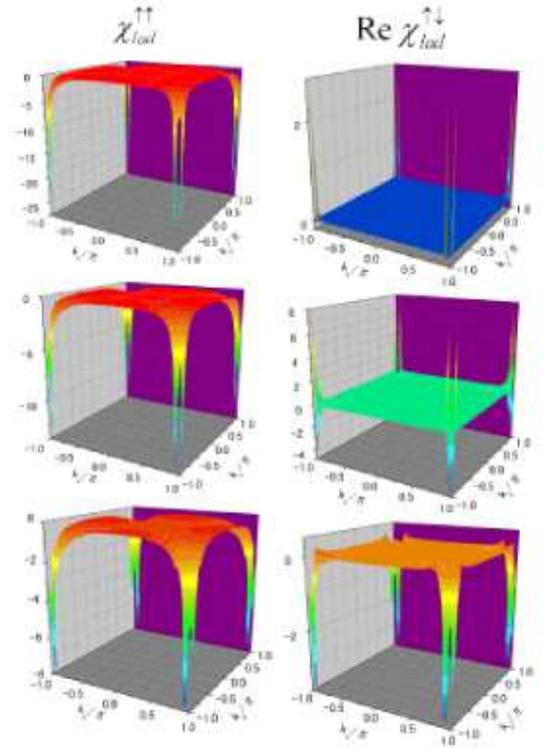}}
\end{center}
\caption{ (color online)  $ \chi _{lad}^{ \uparrow  \uparrow } $ and real part of $ \chi _{lad}^{ \uparrow  \downarrow } $.  We set  $h=0.01$ and $\lambda=0$ in the upper figures, $h=0.1$ and $\lambda=0$ in the middle figures, and $h=0.1$ and $\lambda=0.3$ in the lower figures.}
\label{f11}
\end{figure}

\begin{figure}[tb]
\begin{center}
\scalebox{0.4}{
\includegraphics[width=14.0cm,clip]{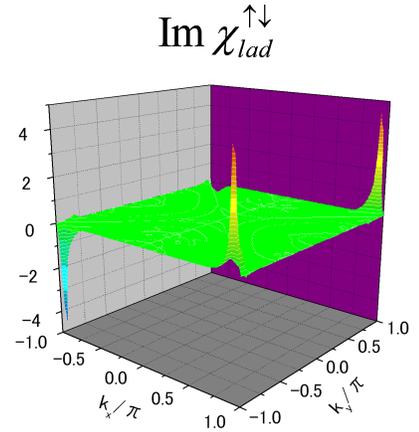}}
\end{center}
\caption{ (color online) Imaginary part of $ \chi _{lad}^{ \uparrow  \downarrow } $ at $h=0.1$ and $\lambda=0.3$.}
\label{f12}
\end{figure}

\begin{figure}[tb]
\begin{center}
\scalebox{0.4}{
\includegraphics[width=21.0cm,clip]{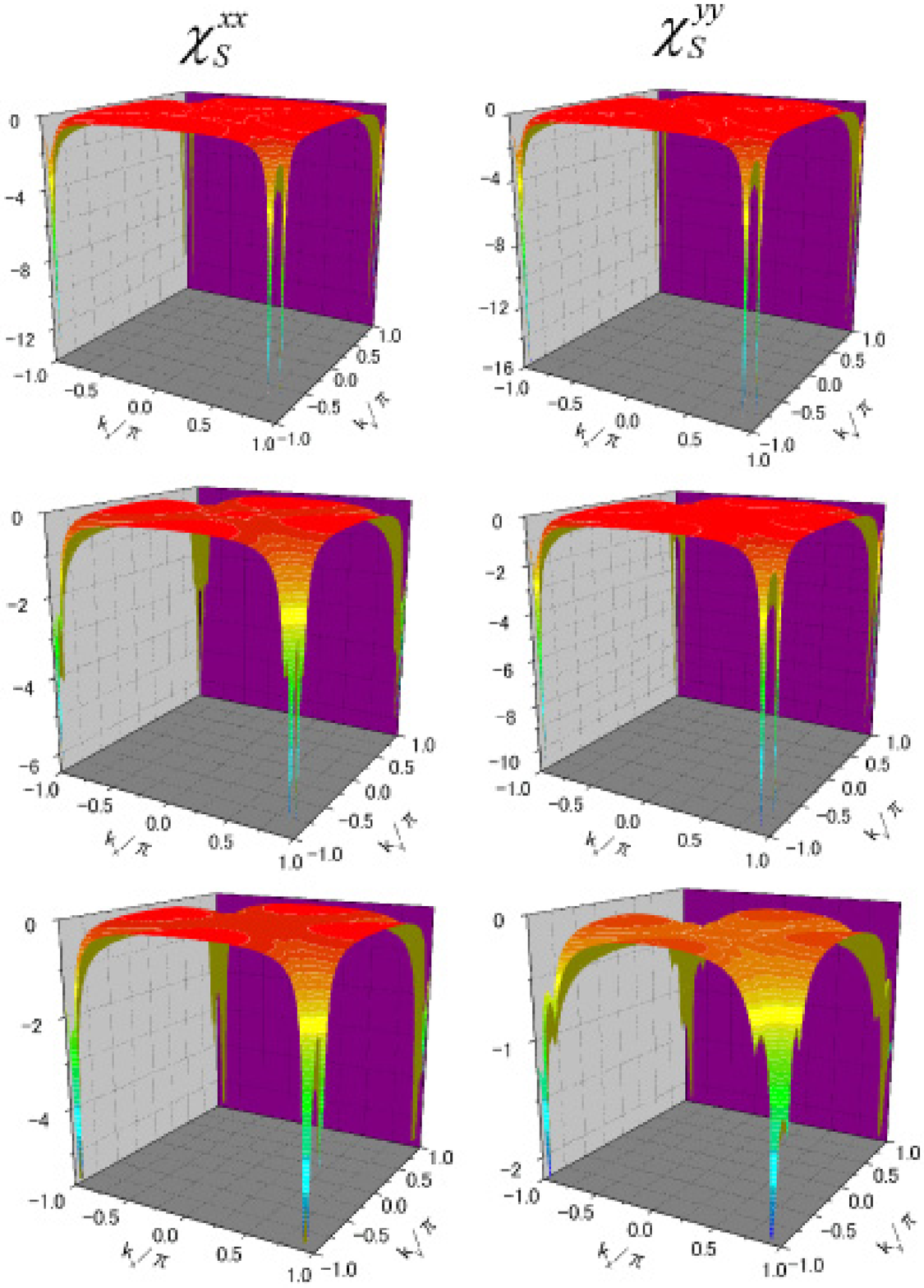}}
\end{center}
\caption{ (color online) $\chi _S^{xx}$ and $\chi _S^{yy}$. We choose 
$h=0.01$ and $\lambda=0$ in the upper figures, $h=0.1$ and $\lambda=0$ in the middle figures, and $h=0.1$ and $\lambda=0.3$ in the lower figures.}
\label{f13}
\end{figure}

Here, we study gap functions which are normalized by the maximum value of $\left| {\Delta _{ s } } \right|$ as a function of wave vector $k$.  Note that  $\Delta _{ \downarrow  \downarrow }  =  - \Delta^* _{ \uparrow  \uparrow } $ and $\Delta _s,\Delta _t \in \reals$ are satisfied due to the weak coupling approximation. It is well known that singlet $d$-wave pairing is dominant over other pairings in the absence of magnetic field and RSOC.\cite{Miyake,Scalapino} 

  Figure \ref{f7} displays real and imaginary parts of the gap function $\Delta _{ \uparrow  \uparrow }$ at $h=0.1$.   We take $\lambda=0$ with $Q_x=8$ and $Q_y=0$ in the upper figure and $\lambda=0.3$, $Q_x=0$ and $Q_y=6$ in the lower figures, where $\Lambda$ takes its maximum value (see Fig. \ref{f2} and Fig. \ref{f3}). As shown in this figure, it has a $f$-wave  symmetry. Note that the imaginary part of $\Delta _{ \uparrow  \uparrow }$ is absent for $\lambda=0$.  We  see that the magnitude of $\Delta _{ \uparrow  \uparrow }$ is enhanced by magnetic field and RSOC. 

We show the real part of triplet gap function $\Delta _t$ in Fig. \ref{f8}. We also take $\lambda=0$, $Q_x=8$ and $Q_y=0$ in the upper figure and $\lambda=0.3$, $Q_x=0$ and $Q_y=6$ in the lower figure. As can be seen from  this figure, it has a $f$-wave like symmetry. Note that in the absence of magnetic field but the presence of the RSOC, $\Delta _t$ vanishes. \cite{Frigeri,Yokoyama3}
We show the real part of singlet gap function $\Delta _s$ in Fig. \ref{f9}. We can find that it has a $d$-wave symmetry. 

We can understand the appearance of $f$-wave symmetry in triplet components by the structures of the spin and charge susceptibilities ($\chi _{S}$ and $\chi _{C}$). They have peaks near $(\pm\pi,\pm\pi)$ as shown in  Fig. \ref{f10}. According to Eq.(\ref{gammauu}), the gap functions tend to have the same sign during the scattering process. Similar discussion is also applicable to   $\chi _{lad}^{ \uparrow  \uparrow }$ and $\chi _{lad}^{ \uparrow  \downarrow }$ (see Figs. \ref{f11} and \ref{f12}), and also Eq.(\ref{gamma}). Therefore, $f$-wave symmetry is favored. We show  typical scattering processes toward $(\pi,\pi)$ by yellow arrows in Fig. \ref{f7} and Fig. \ref{f8}.


Let us explain the origin of the enhancement of the triplet pairing. We  study $\chi _{S}$,  $\chi _{C}$,  $\chi _{lad}^{ \uparrow  \uparrow }$ and  $\chi _{lad}^{ \uparrow  \downarrow }$, which are plotted  in Figs. \ref{f10}- \ref{f12}. \cite{Yokoyama3}
As shown in Fig. \ref{f10}, $\chi _{S}$ is reduced by magnetic field and RSOC while $\chi _{C}$ is almost independent of them. This can be intuitively interpreted as follows.  Magnetic field and spin-orbit coupling cause spin flip process and hence break magnetic fluctuation. On the other hand, spin flip scattering does not affect  charge fluctuation. Therefore,  $\chi _{S}$ depends on  magnetic field and the RSOC while $\chi _{C}$ is almost independent of them. 
Since the positions of the peaks in $\chi _{S}$  and $\chi _{C}$ are almost the same, they compete with each other (see Eq.(\ref{gamma})). The decrease of  $\chi _{S}$ leads to the reduction of singlet $d$-wave pairing and hence triplet pairings could dominate.\cite{Onari}  As shown in Fig. \ref{f11}, $\chi _{lad}^{ \uparrow  \uparrow }$, which contributes to the effective interaction for the heterospin pairings, is also reduced by magnetic field and the RSOC by the same reason, while the real part of $\chi _{lad}^{ \uparrow  \downarrow }$, which contributes to the effective interaction for the equal spin triplet pairings,  is enhanced by magnetic field and the RSOC. This results in the enhancement of the triplet components. Note that the imaginary part of $\chi _{lad}^{ \uparrow  \downarrow }$ emerges at $h=0.1$ and $\lambda=0.3$, although it is negligibly small except for the case in the presence of both of magnetic field and the RSOC. 

Next, we discuss the spin susceptibility for the $x$-direction, $\chi _S^{xx}  = \frac{1}{2}(\chi _{lad}^{ \uparrow  \uparrow }  + {\mathop{\rm Re}\nolimits} \chi _{lad}^{ \uparrow  \downarrow } )$,  and that for the $y$-direction, 
 $\chi _S^{yy}  = \frac{1}{2}(\chi _{lad}^{ \uparrow  \uparrow }  - {\mathop{\rm Re}\nolimits} \chi _{lad}^{ \uparrow  \downarrow } )$. The susceptibility for the $z$-direction $\chi _S^{zz}$ is defined as $\chi _S^{zz}  = \frac{1}{2}\chi _S^{}$. We see that magnetic field strongly suppresses  $\chi _S^{xx}$, while the RSOC strongly suppresses  $\chi _S^{yy}$ and $\chi _S^{zz}$: $\chi _S^{yy}  \sim \chi _S^{zz}  > \chi _S^{xx}$ at $h=0.1$ and $\lambda=0$, and $\chi _S^{xx}  > \chi _S^{yy}  \sim \chi _S^{zz}$   at $h=0.1$ and $\lambda=0.3$,  
 as can be seen from  Fig. \ref{f10} and Fig. \ref{f13}. Since the spin susceptibilities have peaks near $(\pm\pi,\pm\pi)$, the FFLO state coexists with antiferromagnetic fluctuation.

\begin{table}[h]
\caption
{Pairing symmetry and center of mass momentum. Here, ${\bf n}=(0,0,1)$, $\hat{x}$ and $\hat{y}$ are unit vectors oriented  to $x$- and $y$-direction, respectively, and ${\bf h}$ is magnetic field.}
\begin{center}
\begin{tabular}{ccccc}
 \hline
  & \hspace{0.0cm} $h=\lambda=0$   & \hspace{0.0cm} $h=0, \lambda \ne 0$  & \hspace{0.0cm}  $h \ne 0, \lambda=0$ & \hspace{0.0cm} $h \ne 0, \lambda \ne 0$ \\ 
 \hline
 \hline
$\Delta _{ \uparrow  \uparrow }$  & \hspace{0.0cm} =0 & \hspace{0.0cm} $f$-wave  & \hspace{0.0cm} $f$-wave & \hspace{0.0cm} $f$-wave \\
  \hline
$\Delta _{t }$  & \hspace{0.0cm} =0 & \hspace{0.0cm} =0  & \hspace{0.0cm} $f$-wave & \hspace{0.0cm} $f$-wave \\
  \hline
$\Delta _{s }$  & \hspace{0.0cm} $d$-wave & \hspace{0.0cm} $d$-wave  & \hspace{0.0cm} $d$-wave & \hspace{0.0cm} $d$-wave \\
  \hline
$Q$  & \hspace{0.0cm} =0 & \hspace{0.0cm} =0  & \hspace{0.0cm} $\parallel 
 \hat{x},\hat{y}$ & \hspace{0.0cm} $\parallel {\bf n} \times {\bf h}$
 \\
  \hline
\end{tabular}
\end{center}

\label{table1} 
\end{table}

Finally, let us summarize the main results in Table \ref{table1}, including results with $h=\lambda=0$ and  $h=0, \lambda \ne 0$.\cite{Miyake,Scalapino,Yokoyama3} 
The appearance of $d+f$-wave pairing in FFLO state is specific to strongly correlated electron systems (cf Ref.\cite{Matsuo}).

\section{Conclusions}
In this paper, we studied  Fulde-Ferrell-Larkin-Ovchinnikov state of superconductors with and without inversion symmetry, where we used the  Hubbard model on the square lattice near half-filling with the RPA.  We clarified the following points:

1. center of mass momentum 
$Q$ tends to be parallel to $x$- or $y$-axis in the presence of  inversion symmetry, while $Q$ tends to be perpendicular  to the magnetic field in the absence of  inversion symmetry. This can be understood by nesting vectors of the Fermi surfaces. 

2. $d+f$-wave pairing is favored. The hetero spin triplet $f$-wave state is present in the FFLO state unlike state in the superconductors only with the RSOC. The triplet $f$-wave state is enhanced by magnetic field and the RSOC. This stems from the reduction of the spin susceptibilities by the magnetic field and the  RSOC.


%
T. Y. acknowledges support by JSPS. 
This work was supported by
NAREGI Nanoscience Project, the Ministry of Education, Culture,
Sports, Science and Technology, Japan, the Core Research for Evolutional
Science and Technology (CREST) of the Japan Science and Technology
Corporation (JST) and a Grant-in-Aid for the 21st Century COE "Frontiers of
Computational Science" . The computational aspect of this work has been
performed at the Research Center for Computational Science, Okazaki National
Research Institutes and the facilities of the Supercomputer Center,
Institute for Solid State Physics, University of Tokyo and the Computer Center.
%


\end{document}